# Ransomware Detection Using Deep Learning in the SCADA System of Electric Vehicle Charging Station

Manoj Basnet
*Department of Electrical and Computer Engineering*
*University of Memphis*
Memphis, TN, USA

Subash Poudyal
*Department of Computer Science*
*University of Memphis*
Memphis, TN, USA

Mohd. Hasan Ali
*Department of Electrical and Computer Engineering*
*University of Memphis*
Memphis, TN, USA

Dipankar Dasgupta
Department of Computer Science
University of Memphis
Memphis, TN, USA

*Abstract*— The Supervisory control and data acquisition (SCADA) systems have been continuously leveraging the evolution of network architecture, communication protocols, next-generation communication techniques (5G, 6G, Wi-Fi 6), and the internet of things (IoT). However, SCADA system has become the most profitable and alluring target for ransomware attackers. This paper proposes the deep learning-based novel ransomware detection framework in the SCADA controlled electric vehicle charging station (EVCS) with the performance analysis of three deep learning algorithms, namely deep neural network (DNN), 1D convolution neural network (CNN), and long short-term memory (LSTM) recurrent neural network. All three-deep learning-based simulated frameworks achieve around 97% average accuracy (ACC), more than 98% of the average area under the curve (AUC) and an average F1-score under 10-fold stratified cross-validation with an average false alarm rate (FAR) less than 1.88%. Ransomware driven distributed denial of service (DDoS) attack tends to shift the SOC profile by exceeding the SOC control thresholds. The severity has been found to increase as the attack progress and penetration increases. Also, ransomware driven false data injection (FDI) attack has the potential to damage the entire BES or physical system by manipulating the SOC control thresholds. It's a design choice and optimization issue that a deep learning algorithm can deploy based on the tradeoffs between performance metrics.

## I. INTRODUCTION

A smart grid integrates the bidirectional information communication technology (ICT) with multiple stakeholders on top of existing power grids that require highly interoperable, scalable, and intelligent, multilayered architecture [1]. The stakeholders are supervisory control and data acquisition (SCADA), utilities, electric vehicle supply equipment (EVSE), advanced metering infrastructures (AMIs), connected automated/semiautomated electric vehicles (CAEVs), and end-users such as energy prosumers/consumers [2]. The EVSE and EVCS are used interchangeably throughout the paper. Any compromise in confidentiality, integrity, and availability (CIA) of the network resources would result in catastrophic consequences since multiple stakeholders suffer cyber-attacks [3].

Ransomware is one of the rising cyber threats of the modern-day to all the ICT industries and cyber-physical entities. Ransomware encrypts the victim computer's critical files (deployed through some social engineering) to lock them by using symmetric, asymmetric, or hybrid keying. Locky, scareware, and crypto-ransomware are reported in the literature [4], [5]. Static, dynamic, or a combination of both are used for the ransomware analysis [4]. Various machine learning (ML) based techniques such as Logistic regression, Support vector machine (SVM), Decision tree, random forest, deep learning-based approaches are discussed [5] [6]. The Deep learning techniques attempt to discover the complex data structures' unique spatio-temporal patterns by tuning its internal parameters with backpropagation [7]. The weights are the parameters that will be optimized by using variants of the stochastic gradient descent algorithm. DNN, CNN, and LSTM are the most widely used deep learning algorithm.

No substantial work has been done in ransomware detection in the SCADA controlled EVCS as far as this authors' knowledge. The malware attacks in industrial control systems are more heard of but rare for the ransomware family though the systems are vulnerable to ransomware attacks too. Recently, news [8]–[10] have emerged out about ransomware attacks on industrial control system (ICS) targeting the big corporate houses, but risk assessment has not been published yet. The current literature review shows that the study and research are limited to a ransomware attack on industrial control systems or SCADA based systems. The works [11], [12] analyzed the potential ransomware threats and their possible impact on SCADA controlled physical system; however, they lacked an attack simulation and detection framework. This lays the foundation for further research, which involves extensive study, analysis, and experimentation. With this background, we have stepped forward to propose a detection framework.

This paper attempts to quantify the impact of ransomware on EVSE charging. We have simulated the experiments using MATLAB, Simulink, and Python. The main contributions of this paper can be summarized as follows.

 a) A novel, scalable, and interoperable deep learning-based ransomware detection framework has been proposed that could be implemented in multiple vulnerable attack points in SCADA controlled EVCS.

 b) The communication between proposed detection frameworks at various locations has been established so that once ransomware is detected at any point, all other detection frameworks (DF) share the information.

## II. SCADA CONTROLLED EVCS

SCADA facilitates the management of remote access to real-time data and channels. It issues automated or operator driven supervisory commands to remote stations (field devices) [13]. The underlying control system of most critical infrastructures such as power, energy, water, manufacturing plants, traffic lights, nuclear plants is SCADA [14]. SCADA consists of sensors, Programmable Logic Controllers (PLC), actuators, Remote Terminal Units (RTU), supervisory station, backend server, human-machine interface communication link. It needs constant vigilance of the target physical plant through the communication link. Sensors are primarily used for data acquisition purposes in the plant's physical environment or process being monitored. These are embedded in the various dynamics of the process.

The Fig.1 represents the high-level block diagram of remote SCADA communicating/controlling the multiple EVSE through 5G communication infrastructure. It also depicts the threat actors in the system. The SCADA is constantly monitoring the state of charge (SOC) of the battery energy storage (BES) at EVCS to control the charging.

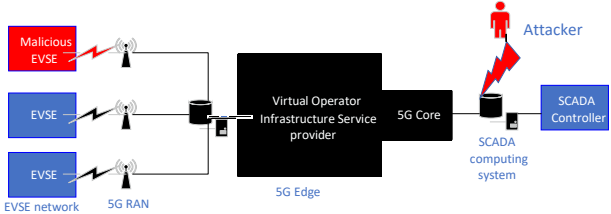

Fig. 1. SCADA system connected to EVSE through 5G ( controlled)

## III. RANSOMWARE ATTACK MODELING ON EVSE

Criminally motivated organizations such as WIZARD SPIDER runs ransomware-as-a-service(RaaS) to target the big game hunting [15]. An attacker encrypts the critical files: access control files, historian data, input scripts, the communication packet at SCADA. The attack motif may not necessarily be a full shutdown of the system but might interrupt normal operations. An operator may not pass the supervisory command or ride over command to the field devices at EVCS. While the victim pays the ransom, it is not guaranteed that the hacker would give a decryption key. Once the attacker gains access to the cyber-physical system's critical files, there are two ways they can harm the operations first by initiating a DDoS attack and second by FDI attack. As reported in recent attacks [16]–[18], there are relatively easy attacks to start. The severity of ransomware attacks to the critical infrastructures depends on how long they captivate the network and how much false data they can inject.

We simulate the system in MATLAB Simulink, where the remote SCADA controls the charge and discharge of the battery energy storage (BES) of the EVSE by issuing charge/discharge command to the ideal switches. The control commands are based on the SOC of the BES. The control is designed so that the BES should discharge, i.e., charge to EV if SOC>80 % or charge from PV or DC source if SOC <35 % between the timeframe of 50 seconds to 150 seconds. The ransomware triggered DDoS on SCADA generally imposes a delay in the control commands to/from the EVSE. We have simulated the effect of ransomware induced delay ranging from zero to five minutes in the charging behavior. Secondly, we have simulated the effect of ransomware triggered FDI attack by manipulating SOC thresholds set to control charge/discharge.

## IV. PROPOSED FRAMEWORK FOR RANSOMWARE DETECTION

The prime target for ransomware could be any field devices (such as PLC, RTU, IoT devices for process dynamics control, and data acquisition), control, and monitoring systems such as supervisory stations and HMI. Reconnaissance of the vulnerabilities in the field devices might need domain knowledge and configuration in the air-gapped system. Also, poorly developed protocols in field devices have easily exploitable authentication, authorization, and access control issues [19]. However, internet-facing field devices could be easily scanned and exploited using available scanning tools such as Shodan, ZoomEye, Censys. The easy target might be HMI or supervisory station because they are the one who controls and monitors all field devices. That's why ransomware attacks in these computing systems that access the SCADA backend could be dangerous. The ease of attack comes here as these components could be treated more or less like the IT system. Besides that, the attacker might not need in-depth design and domain knowledge, unlike in PLC and RTU. We propose the novel ransomware monitoring and detection system in Fig. 2. It continuously monitors and detects the ransomware attack in SCADA, power generation/transmission and distribution network, EVSE network, and CAEV. The unit ransomware detection framework (RDF) architecture is presented in Fig. 3 for the SCADA. Likewise, RDF could be implemented for all remaining three layers beneath SCADA.

The monitoring and detection system has two phases offline and online. In the offline phase, the ransomware samples and the benign samples are collected from the attack. Important features are extracted from the collected samples by using assembly frequency analysis. Finally, the feature database is built to train and validate the deep learning-based model. Training is generally the curve fitting process based on weight optimization. Simultaneously, validation is the way to tune the architectures and hyperparameters to get the best model before deploying it. Deep learning does not provide a sophisticated way to automate the process of tuning the model. In the online phase, the feature from the real-time traffic is extracted using the same method as in training and is fed to the model to detect whether it is ransomware or a benign file. Once the ransomware is detected, an appropriate control strategy is activated to turn the backups on, isolate the physical layer; or shut down the system in the worst-case scenario.

### A. Datasets

We have collected 561 ransomware samples and 447 normal samples summing up to 1008. Ransomware binaries were collected from VirusTotal, whereas normal samples were extracted from the Windows operating system. The normal samples have similar sizes as ransomware (50KB – 8 MB), and files with cryptographic behavior (Filezilla, Winscp, OpenSSH, and so on) are also included. The ransomware considered here is Crypto ransomware. Feature selection is made by using frequency analysis of assembly instructions. Individual assembly instruction, including the grouping count, was taken.

Assembly Grouping is grouped as Data transfer, Arithmetic, Logical, shift, and so on. A dynamic binary instrumentation technique was used to extract assemblies. It's a dynamic running of malware samples in a controlled virtual unit via the PIN framework. Further feature extraction was done extensively using custom python programming (Grouping, frequency generation, unique features, and CSV file generation).

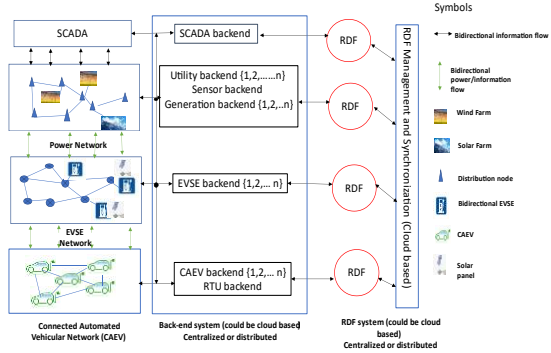

Fig. 2. Proposed ransomware detection system for smart grid architecture.

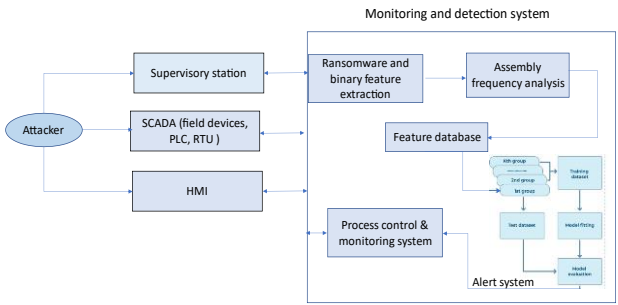

Fig.3. The internal architecture of the proposed RDF monitoring and detection system

### B. Deep Learning Architectures and parameters setting for simulations

A three-layered DNN with two hidden layers, each layer with 64 hidden neurons for the binary classification, is implemented for the ransomware detection in SCADA. Apart from that, the hidden layer uses the ReLu activation function since it has better convergence property and prevents the sigmoid function's problem, which tends to produce vanishing and exploding gradients. The defacto standard for the optimizer, Adam, is implemented in the DNN. The default activation function in the output layer and the loss are sigmoid and binary cross-entropy, respectively, for binary classification. The proposed model implements the L1-L2 regularizers. These regularizers apply penalties on layer parameters or layer activity during optimization. These penalties are incorporated in the loss function that the network optimizes.

LSTM is the variant of the recurrent neural network (RNN) developed to eliminate the vanishing gradient problem of RNN and is significantly more complex than traditional neural units [20]. LTSM Cell Architecture: Each cell has four sets of weights that feed into it (instead of one). Output squashing can take any activation function we want. It learns 1. What/when to let something in, 2. When to forget, 3. What/when to let something out. Most of the architecture is similar to that of DNN shown [21] except the cell structure. A 140, 64, 64, 64, 1 architecture is used for binary classification with neuron dropout of 10 % between each hidden layer. The architecture mentioned above is read as # of input units=140, # of LSTM cells in first hidden layer= 64, # of LSTM cells in second hidden layer= 64, # of LSTM cell in third hidden layer= 64, and # of output units=1. The first and last layers are the input and output layers with corresponding nodes, while the middle layers represent the hidden layers with corresponding nodes.

1D CNN is the variant of the CNN designed to convolve 1D data vectors rather than convolving 2D or higher. Our model consists of two convolution layers with a kernel size of 64 and filter length 3, each followed by a max-pooling layer. A fully connected hidden layer with 128 hidden units is between the max-pooling and output layers, with a unit dropout rate of 50%.

### V. SIMULATION RESULTS AND DISCUSSION

#### A. Simulation Conditions and Process

In this work, all the simulations and coding for ransomware detection are created in Python 3.7.4 in the Jupyter lab (version 1.1.4) under the free and open-source Anaconda distribution. Intel® Core™ i5-3470 @ 3.20 GHz processor with 8.00 GB RAM and 64 -bit Windows 10 OS is used in the experiment. Before starting the simulation, data preprocessing has been done to scale down diverse features with a different magnitude between 0 and 1 using the Keras standard preprocessing. scale() library. Also, the categorical output classes ransomware vs. normal are binarized to 0 and 1, respectively. Out of 1008 samples, 30% of data are preserved for testing, the other 30% are preserved for validation; and finally, the model is trained with the remaining 40 % of the samples for a single run of the experiment. Each of these train, test, and validation category was mutually exclusive. Moreover, the training and validation are done with 10-fold stratified cross-validation to check the model consistency and reproducibility. Our experiment is done with a batch size of 100 with 70 epochs for all deep learning algorithms.

With ransomware attackers having access to the critical process file of SCADA, they can hijack the system for the time they want, and they can inject or manipulate the control variables. These impacts are observed under the ransomware driven DDoS attack trying to deprive the legitimate users and FDI attack to decimate the EV charging.

#### B. Ransomware driven DDoS attack

As mentioned in section III, the charging behavior of BES is observed as the DDoS attack penetration increases from no delay to delay up to five minutes. The five SOC transition edges are recorded with a tuple of <SOC, time> at those edges between control action period of 50 seconds to 150 seconds, as shown in figure 4. The SOC0 represents the SOC behavior without any attack, i.e., normal behavior. SOCi represents the SOC behavior with i minutes of delay due to DDoS. With further data analysis as the SOC0 as a reference signal, the state transition due to the attack is delayed by a minimum of 0.139% to a maximum of

4.84%. This also forces SOC to exceeds the control thresholds, potentially impacting the battery dynamics.

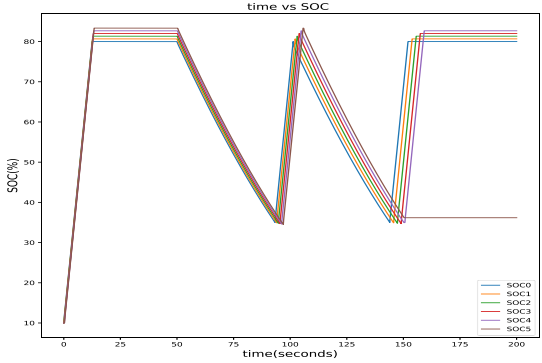

Fig. 3. Ransomware driven DDoS attack impact on charging behavior

The severity in SOC threshold crossing increases as the attack penetration increases with a minimum of 0.29% to a maximum of -54.73% compared to the normal operating SOC. With five minutes or more delay induced by DDoS penetration, the charging profile becomes so worse that it just stays above the lower threshold of 35% SOC without charging ever again after 150 seconds. Therefore, it can be concluded that even the simple DDoS driven by ransomware could produce erroneous control commands, which is enough to detriment the EV charging.

### C. Ransomware driven FDI attack

Similar architecture has been implemented to model the FDI attack, except ransomware attacker manipulates the SOC thresholds to make control decisions at state transition diagram of SCADA. The Fig. 5 represents the different SOC profiles S0C0, SOC1, SOC2, SOC3 with respective state transition thresholds tuples of (35,80), (10,90), (5,95), and (0,100). The tuple has a minimum and maximum value of SOC that should not be exceeded and issue either charging or discharging command at the instant of crossing them. The BES at EVSE has specific energy density, power density with a limited number of cycles. The FDI attack can abruptly change the charging behavior and damage the BES or physical system in the worst-case scenario, as depicted in the figure below.

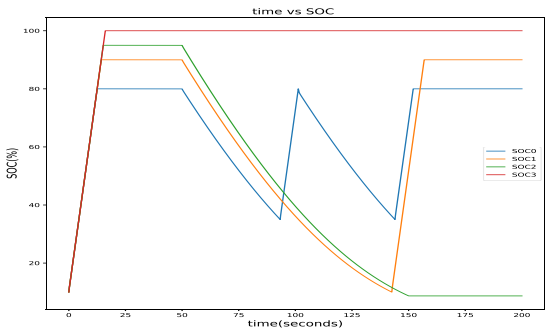

Fig. 4. Ransomware driven FDI attack impact on charging behavior

### D. Deep learning-based analysis

As shown in Fig.6, all models achieved at least 98 % accuracy, with DNN being very smooth, representing the perfect fitting during the training phase. 1D CNN has a lot of noise after ten epochs. Though LSTM has shown small fluctuation initially for the few epochs, it is the most smooth algorithm that achieves the desired accuracy within ten-eleven epochs. Therefore, one can conclude, finely tuned LSTM looks superior for one experiment, though model replicability is the issue of all deep learning techniques. Let's see the results of 10 stratified cross-validations in Table I. The area under the curve (AUC) refers to the degree of separability of the model to distinguish between different classes. LSTM seems like the right option in terms of AUC because of its highest AUC and the least standard deviation (std). However, CNN seems superior in terms of accuracy.

TABLE I. THE AREA UNDER THE CURVE (AUC) AND ACCURACY (ACC) OF 10 FOLD STRATIFIED CROSS-VALIDATION

| DL methods | AUC(Mean) | AUC (Std) | ACC (mean) | ACC (std) |
|---|---|---|---|---|
| DNN | 98.17 % | 2.35 % | 98.30 % | 1.52 % |
| CNN | 98.16 % | 1.10 % | 98.73 % | 1.27 % |
| LSTM | 98.94 % | 0.72 % | 97.59 % | 1.91 % |

The Training of DNN is found to be way faster than at least five times faster than LSTM and sixteen times faster than CNN for 70 epochs, as shown in Table II. It is because of the least units and parameters used in DNN in comparison to CNN and LSTM. However, As evident from Fig. 6, the desired accuracy could be achieved using fewer epochs than 70, reducing LSTM and CNN's training time. Table. III shows the mean precision, recall, f1-score, and FAR with standard deviation all in % after a 10 fold stratified cross-validation. The CNN model achieves the best f1-score with minimum FAR.

TABLE II. TRAINING TIME

| DL methods | Training time |
|---|---|
| DNN | 1.349 seconds |
| CNN | 16.581 seconds |
| LSTM | 5.98 seconds |

TABLE I. PERFORMANCE METRICS AFTER 10 FOLD CROSS-VALIDATION

| DL methods | Precision Mean \| std in % | Recall Mean \| std in % | F1-score Mean\| std in % | FAR Mean\| std in % |
|---|---|---|---|---|
| DNN | 98.45 \| 1.71 | 97.92 \| 1.94 | 98.17 \|1.46 | 1.88 \|2.07 |
| CNN | 99 \| 1.22 | 97.33 \| 1.77 | 98.35 \|1.14 | 1.30 \|1.59 |
| LSTM | 98.70 \|1.75 | 97.66 \| 2.17 | 98.16 \|1.47 | 1.56 \|2.10 |

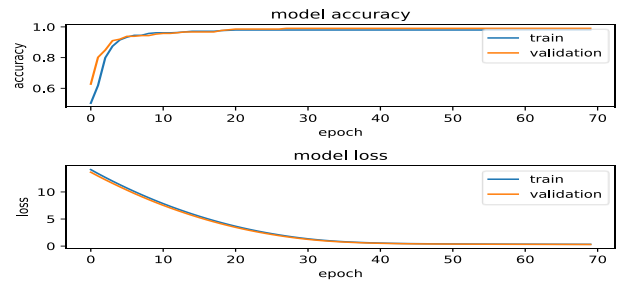

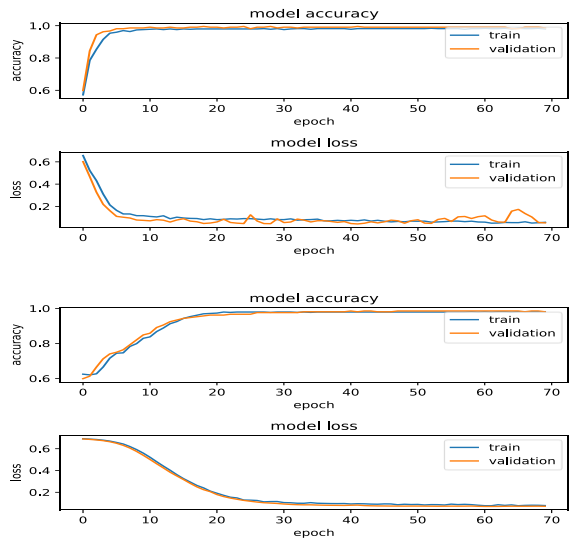

Fig. 5. Model accuracy vs. loss for a single experiment for DNN, 1D CNN, and LSTM top to bottom.

## VI. CONCLUSION

This paper successfully implemented the ransomware attack in a simulated scenario of SCADA controlled EVSE, analyzed the impact on charging behavior of BES and proposed/assessed the novel deep learning-based ransomware detection framework. The findings are concluded as follows.

- Ransomware driven DDoS attack tends to shift the SOC profile by exceeding the SOC control thresholds. The severity has been found to increase as the attack progress and penetration increases.
- Ransomware driven FDI attack has the potential to damage the entire BES or physical system by manipulating the SOC control thresholds.
- DNN could be a choice in the SCADA with a minimal computational time for simplicity and less time complexity.
- In terms of AUC, LSTM is the best option with the least variance and maximum mean AUC.
- If high accuracy is needed, then 1D CNN could be implemented with the cost of model complexity.

SCADA system needs the utmost cyber-physical security with minimal average FAR. The proposed model ensures less than 1.88% of the average FAR, with CNN being the best. Therefore, it is a design choice which DL algorithms to deploy based on ACC, f1-score, AUC, model complexity. Our future work will include the impact analysis of ransomware on a sophisticated smart charging algorithm.